\documentclass[twoside,twocolumn,9pt]{article}
\usepackage{extsizes}
\usepackage[super,sort&compress,comma]{natbib} 
\usepackage[version=3]{mhchem}
\usepackage[left=1.5cm, right=1.5cm, top=1.785cm, bottom=2.0cm]{geometry}
\usepackage{balance}
\usepackage{times,mathptmx}
\usepackage{sectsty}
\usepackage{graphicx} 
\usepackage{lastpage}
\usepackage{amsmath}
\usepackage{amssymb}
\usepackage[format=plain,justification=justified,singlelinecheck=false,font={stretch=1.125,small,sf},labelfont=bf,labelsep=space]{caption}
\usepackage{float}
\usepackage{fancyhdr}
\usepackage{fnpos}
\usepackage[english]{babel}
\addto{\captionsenglish}{%
  
}
\usepackage{braket}
\usepackage{array}
\usepackage{droidsans}
\usepackage{charter}
\usepackage[T1]{fontenc}
\usepackage[usenames,dvipsnames]{xcolor}
\usepackage{setspace}
\usepackage[compact]{titlesec}
\usepackage{hyperref}
\usepackage{bm}
\usepackage{dcolumn}

\usepackage{epstopdf}

\definecolor{cream}{RGB}{222,217,201}

\begin{document}

\pagestyle{fancy}
\thispagestyle{plain}
\fancypagestyle{plain}{

\renewcommand{\headrulewidth}{0pt}
}

\makeFNbottom
\makeatletter
\renewcommand\LARGE{\@setfontsize\LARGE{15pt}{17}}
\renewcommand\Large{\@setfontsize\Large{12pt}{14}}
\renewcommand\large{\@setfontsize\large{10pt}{12}}
\renewcommand\footnotesize{\@setfontsize\footnotesize{7pt}{10}}
\makeatother

\renewcommand{\thefootnote}{\fnsymbol{footnote}}
\renewcommand\footnoterule{\vspace*{1pt}%
\color{cream}\hrule width 3.5in height 0.4pt \color{black}\vspace*{5pt}} 
\setcounter{secnumdepth}{5}

\makeatletter 
\renewcommand\@biblabel[1]{#1}            
\renewcommand\@makefntext[1]%
{\noindent\makebox[0pt][r]{\@thefnmark\,}#1}
\makeatother 
\renewcommand{\figurename}{\small{Fig.}~}
\sectionfont{\sffamily\Large}
\subsectionfont{\normalsize}
\subsubsectionfont{\bf}
\setstretch{1.125} 
\setlength{\skip\footins}{0.8cm}
\setlength{\footnotesep}{0.25cm}
\setlength{\jot}{10pt}
\titlespacing*{\section}{0pt}{4pt}{4pt}
\titlespacing*{\subsection}{0pt}{15pt}{1pt}

\fancyfoot{}
\fancyfoot[LO,RE]{\vspace{-7.1pt}\includegraphics[height=9pt]{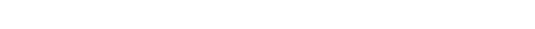}}
\fancyfoot[CO]{\vspace{-7.1pt}\hspace{13.2cm}\includegraphics{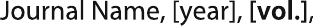}}
\fancyfoot[CE]{\vspace{-7.2pt}\hspace{-14.2cm}\includegraphics{head_foot/RF}}
\fancyfoot[RO]{\footnotesize{\sffamily{1--\pageref{LastPage} ~\textbar  \hspace{2pt}\thepage}}}
\fancyfoot[LE]{\footnotesize{\sffamily{\thepage~\textbar\hspace{3.45cm} 1--\pageref{LastPage}}}}
\fancyhead{}
\renewcommand{\headrulewidth}{0pt} 
\renewcommand{\footrulewidth}{0pt}
\setlength{\arrayrulewidth}{1pt}
\setlength{\columnsep}{6.5mm}
\setlength\bibsep{1pt}

\makeatletter 
\newlength{\figrulesep} 
\setlength{\figrulesep}{0.5\textfloatsep} 

\newcommand{\topfigrule}{\vspace*{-1pt}%
\noindent{\color{cream}\rule[-\figrulesep]{\columnwidth}{1.5pt}} }

\newcommand{\botfigrule}{\vspace*{-2pt}%
\noindent{\color{cream}\rule[\figrulesep]{\columnwidth}{1.5pt}} }

\newcommand{\dblfigrule}{\vspace*{-1pt}%
\noindent{\color{cream}\rule[-\figrulesep]{\textwidth}{1.5pt}} }

\makeatother

\newcommand{\orcid}[1]{\href{https://orcid.org/#1}{\includegraphics[width=8pt]{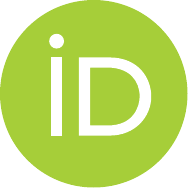}}}
\newcommand*\mycommand[1]{\texttt{\emph{#1}}}
\newcommand{\fp}{\it first-principles}
\newcommand{\avg}[1]{\langle #1 \rangle}
\newcommand{\red}[1]{{\color{red} #1}}
\newcommand{\blue}[1]{{\color{blue} #1}}

\twocolumn[
  \begin{@twocolumnfalse}
  {\includegraphics[height=30pt]{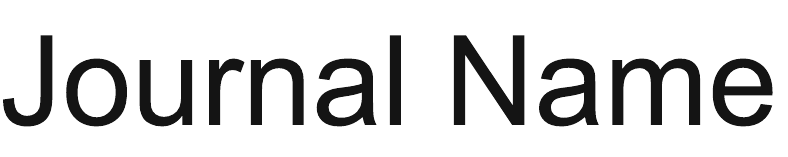}\hfill%
 \raisebox{0pt}[0pt][0pt]{\includegraphics[height=55pt]{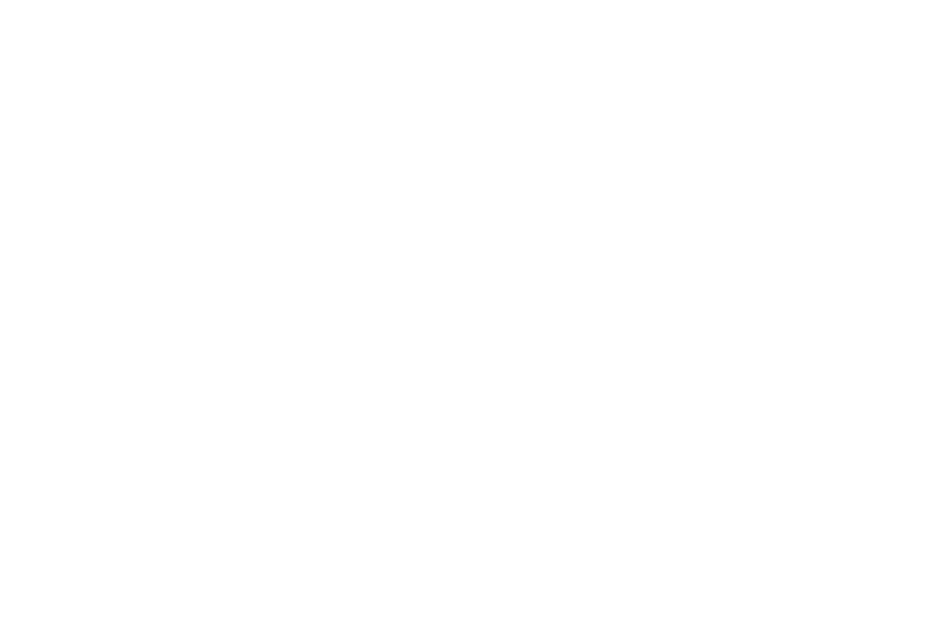}}%
 \\[1ex]%
 \includegraphics[width=18.5cm]{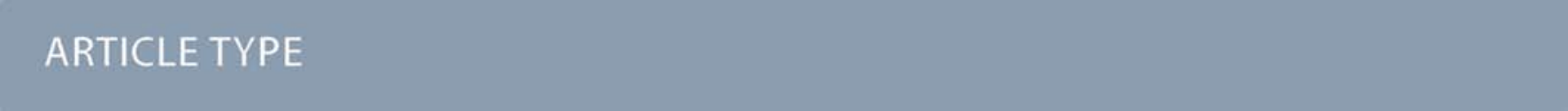}}\par
\vspace{1em}
\sffamily
\begin{tabular}{m{4.5cm} p{13.5cm} }

\includegraphics{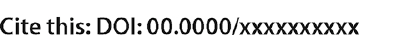} & \noindent\LARGE{\textbf{Bandgap evolution in nanographene assemblies}} \\
\vspace{0.3cm} & \vspace{0.3cm} \\

 & \noindent\large{F. Crasto de Lima\textit{$^{a}$} \orcid{0000-0002-2937-2620} and A. Fazzio\textit{$^{a}$} \orcid{0000-0001-5384-7676}} \\

\includegraphics{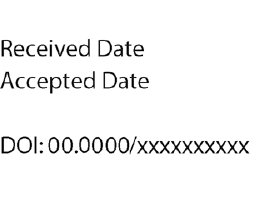} & \noindent\normalsize{ Recently cycloarene has been experimentally obtained in a self-assembled structure, forming graphene-like monoatomic layered systems. Here, we establish the bandgap engineering/prediction in cycloarene assemblies within a combination of density functional theory and tight-binding Hamiltonians. Our results show that the inter-molecule bond density rules the bandgap. The increase in such bond density increases the valence/conduction bandwidth decreasing the energy gap linearly. We derived an effective model that allows the interpretation of the arising energy gap for general particle-hole symmetric molecular arranges based on inter-molecular bond strength.} \\

\end{tabular}

 \end{@twocolumnfalse} \vspace{0.6cm}

  ]

\renewcommand*\rmdefault{bch}\normalfont\upshape
\rmfamily
\section*{}
\vspace{-1cm}


\footnotetext{\textit{$^{a}$~Brazilian Nanotechnology National Laboratory (LNNano), Brazilian Center for Research in Energy and Materials (CNPEM), 13083-970, Campinas, SP, Brazil.}}




\section{Introduction}

The ability to predict and control the energy gap in 2D materials allows the rational design of nanodevices. Graphene, the first discovered 2D material \cite{SCIENCEnovoselov2004}, has emerged as a promise to different phenomena and device applications \cite{NATMATgeim2007, NATNANOschwierz2010, NATNANOsurwade2015, NATELEClin2019}. Its monoatomic layer has shown to be very resistant against tearing while retaining the flexibility inherent in 2D systems \cite{PMSdimitrios2017}. Together with these mechanical properties, the emerging Dirac cone makes its electronic mobility reach $10^5$\,cm$^2$V$^{-1}$s$^{-1}$ \cite{SSCbolotin2008, SCIENCEbanszerus2015}. The 2D structure, combined with interesting electronic properties, lead to the emerging field of vdW heterostructures of 2D systems for device designs. Graphene functionalization and interfacing have also brought a degree of freedom in tunning its properties \cite{PRLpadilha2015, PCCPdelima2018, CARBONpadilha2019}. For instance, upon transition metal absorption/interfacing, graphene presents magnetic \cite{JPCCzanella2008, PRBlima2011, MCPdeLima2017, NATPHYbenitez2018} and topological \cite{PRBmin2006, PRBacosta2014, PRBkochan2017, PRLhogl2020} phases ruled by the spin-orbit coupling. One lasting challenge in graphene systems is creating semiconducting phases with a wider gap to expand its application in nanoelectronics and photonics. However, finding ways to fine-tune the graphene energy gap around $0.5$\,eV, and higher is still an open problem. Inducing an energy gap in graphene has been proposed by applying an external field in graphene bilayers \cite{SCIohta2006, NATzhang2009, JPCMpadilha2012, RSCADVtang2017}. Additionally, a small gap ($\sim 100$\,meV) can be opened by interfacing graphene with boron-nitrite \cite{PRBgiovannetti2007, NATCOMMjung2015} and other substrates \cite{NATMATzhou2007, PRLvarchon2007, PRLnevius2015}. Nano-technological applications could profit from a graphene system with an energy gap of the silicon order.

Recently nanoporous graphene has emerged as a solution to higher gap graphene (with $\sim 1$\,eV gap) \cite{SCIENCEmoreno2018, CARBONbohayra2019}. Such systems can be viewed as an assembly of carbon molecules forming graphene-like systems, usually presenting porous or distortions. Given the interest in those systems, they have suffered a rapid development with many molecule/structures synthesized \cite{CSRnarita2015, CARBONconchi2020}. The variety of such molecular systems allows for many possible self-assembled structures, leading to rich nanoporous graphene structures. For instance, different coronoids structures were synthesized on Au(111) \cite{JACSdi2020}, while cellulose acetates have shown to evolve to graphene monolayers on Cu surface \cite{CARBONmingguang2021}. In recent work, Qitang et al. have identified the self-assembly of cycloarenes in Au(111) surface \cite{JACSfan2020}. Such graphene frameworks can host a semiconducting phase, envisioned for graphene electronics devices. Additionally, such structures can preserve graphene mechanical characteristics: (i) strong planar sp2 bond; (ii) weak interlayer vdW interaction for future heterostructures devices. Moreover, the nanoporous structure allows for further chemical functionalization \cite{CSRshun2016}.

In this paper, we correlate the cycloarene assemblies with their inherent energy gap. Here we show that the energy gap depends linearly on the inter-molecule bond density, being almost independent of the assembly structure. By using a density functional theory (DFT) fitted tight-binding (TB) model, we have tracked different assemblies' electronic structure. We propose an effective model that allow to understand the gap variation phenomena while showing its generality to other molecular-based porous graphenes.

\begin{figure*}[t]
\includegraphics[width=2\columnwidth]{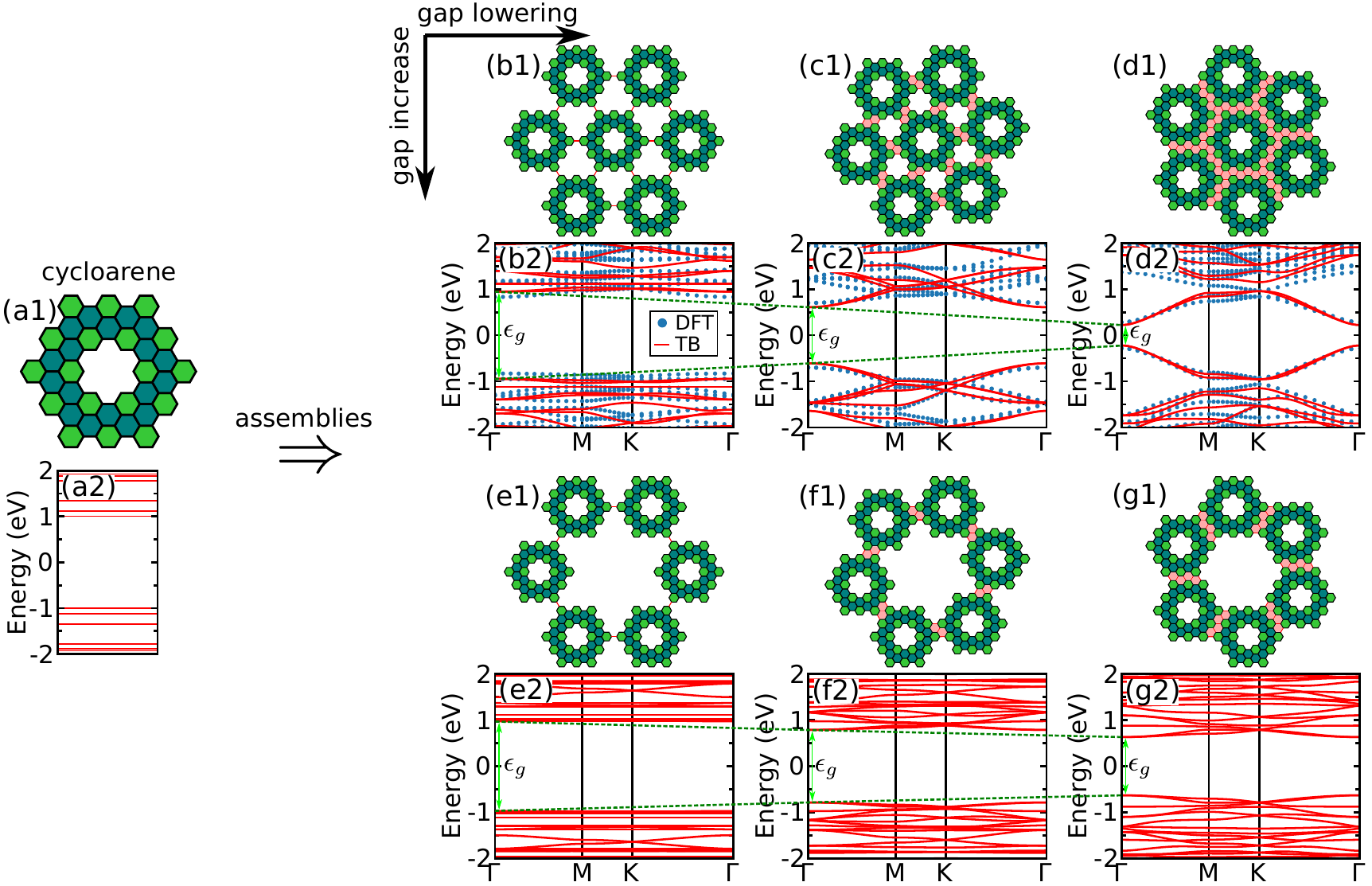}
\caption{\label{2d-assembly} (a1) Atomic configuration of cycloarene and (a2) its energy spectra. (b1)-(g1) 2D assemblies of cycloarene, the shaded red region marks the bonds between neighboring cycloarene molecules. (b2)-(g2) Respective band structures of the 2D assemblies, the green dashed line highlights the energy gap lowering with increased intermolecular bonds.}
\end{figure*}

\section{Methods}
We have performed DFT calculations as implemented in the Quantum Espresso code \cite{JPCMgiannozzi2009}. The exchange-correlation was included within the PBE functional \cite{PRLperdew1996, PRLperdew1997} in a plane-wave base with a cutoff energy of $30$\,Ry. We also performed HSE06 hybrid-functional calculations \cite{JCPheyd2003, JCPkrukau2006} to access a better estimation of the cycloarene energy gap. For the isolated molecule, a single k-point was considered for Brillouin zone integration, while for the assemblies, a k-mesh of $3\times3\times1$ special points \cite{PRBmonkhorst1976}. The electron-ion interaction was described within the projected augmented wave method (PAW) \cite{PRBblochl1994}, where all atoms were relaxed until the forces were lower than $10^{-3}$ (Ry/a.u.). 

In order to construct the assemblies, we are taking the cycloarene molecule as a building-block. Cycloarene is a coronoid nano-graphene, C108-molecule, as shown in Fig.~\ref{2d-assembly}(a). We have calculated an isolated molecule with the DFT approach, presenting a carbon-carbon mean distance of $1.43$\,{\AA} and a non-magnetic ground state ruled by the molecule's armchair termination. {The non-magnetic ground state is $0.92$\,eV lower in energy then a doublet configuration (1 unpaired electron).} Our converged isolated molecule follows the same carbon-carbon distance distribution observed experimentally \cite{JACSfan2020}. Our goal is to explore assemblies of C108 nano-graphenes, which have a large number of carbon atoms in the periodic cell.

To expand our DFT calculations for higher number of atoms, we derived a nearest-neighbors (NN) TB model. In this model, we considered the on-site energy of the carbons $p_z$ orbitals as zero. The hamiltonian is described as 
\begin{equation}
H=\sum_{\langle i,j \rangle} t\,|i \rangle \langle j |,
\end{equation}
with the summation over NN sites. By taking $t=2.7$\,eV, the TB model could predict the DFT bandgap energy which we validate for three different assemblies band structure [Fig.~\ref{2d-assembly}(b)-(d)]. The TB model deviates from the DFT bandgap on average by only $0.08$\,eV. Such small variation, compared with the energy gap scale ($\sim 1$\,eV) does not alter the discussion presented below.

\section{Results and Discussion}

The isolated cycloarene molecule presents a calculated energy gap of $2.01$\,eV, on the PBE functional level. By assembling it in different structures, we found a gap reduction. The control of molecules' position on surfaces, by atomic force microscopy and scanning tunneling microscopy tips, and nano-patterning in graphene allows the precise generation of ordered structures \cite{NANOSCAfeng2012, RPPsaw2014, CPCsun2014, NATREVCHEMpavlicek2017, NLcrasto2020, JPCMqiushi2021}. Additionally, by varying experimental growth degrees of freedom, highly ordered structures can self-assemble \cite{ACSNANOfeng2019, CHEMRXIVinayeh2020}. Here we explore different cycloarene assemblies by varying the number of bonds between neighboring molecules and the lattice geometry. For a hexagonal cell with perfect triangular geometry, only three possible structures can be constructed, with $1$, $3$, and $5$ bonds between adjacent molecules, as shown in Fig.~\ref{2d-assembly}(b), (c), and (d), respectively. Such structures give rise to variations of the energy gap in the assembly, being $1.89$, $1.22$, and $0.44$\,eV  [Fig.~\ref{2d-assembly}(b)-(d)]. Following the same inter-molecule bonding order, we can construct three molecular AB sublattice graphene-like structures, as shown in Fig.~\ref{2d-assembly}(e)-(g). In these graphene-like structures, the inter-molecule bond is equal to the perfect triangular structure, $1$, $3$, or $5$. However, the inter-molecule bond density
\begin{equation}
\eta = n_b/N,
\end{equation}
with $n_b$ inter-molecule bonds and $N$ molecules in the unit cell, gets lower than the triangular systems by a factor of $2$. This lowering of $\eta$ weakens the coupling between neighboring molecules and leads to an increase in the energy gap of $1.85 \rightarrow 1.95$, $1.22 \rightarrow 1.57$, and $0.44 \rightarrow 1.26$\,eV [Fig.~\ref{2d-assembly}.(e)-(g)]. Comparing the band structures of Fig.~\ref{2d-assembly}, we see that the effective inter-molecule interaction, ruled by $\eta$, alter the HOMO/LUMO molecular states dispersion, that is, alter its bandwidth. Such an effect lower the energy gap for high-dispersive bands [Fig.~\ref{2d-assembly}(d2)] while converges to the isolated molecular gap ($2.01$\,eV) for low-dispersive bands [Fig.~\ref{2d-assembly}(e2)]. 

\begin{figure}
\includegraphics[width=0.8\columnwidth]{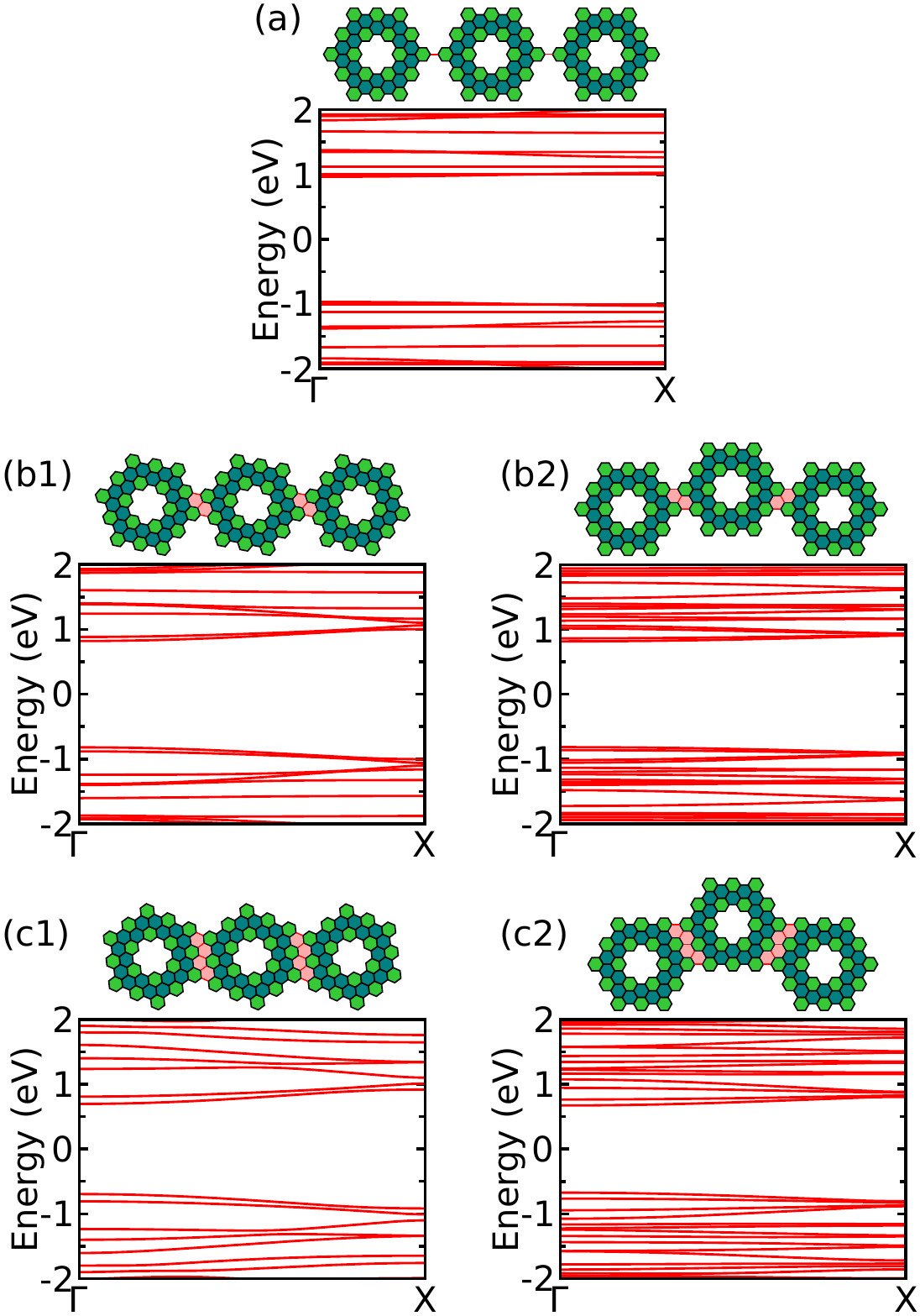}
\caption{\label{1d-assembly} 1D assemblies of cycloarene forming hole decorated graphene nanorribons. The upper panels show the atomic configuration with red shaded regions indicating the inter-molecular bonds. (a) The lower inter-molecular bond density 1D assembly, (b)-(c) the second and third lowest inter-molecular bond density assembly with two configurations: (a1)-(b1) linear, and (b2)-(c2) sinuous.}
\end{figure}

Linear assemblies between cycloarenes are present in its experimental realization \cite{JACSfan2020}. By studying different 1D periodic geometries, we see also a dependence of the energy gap with the different inter-molecular bond density. For a linear array of cycloarene with only one bond between molecules, we see an energy gap of $1.93$\,eV, Fig.~\ref{1d-assembly}(a). Such gap is lowered to $1.64$ and $1.63$\,eV for linear and sinuous molecular arrays with three bonds ($\eta = 3$) between neighboring molecules [Fig.~\ref{1d-assembly}(b1)-(b2)]; and to $1.39$ and $1.35$\,eV for the five bonds case [Fig.~\ref{1d-assembly}(c1)-(c2)]. Here we can understand that (i) the gap has a weak dependence on the system structure (linear or sinuous), instead (ii) the inter-molecule bond density rules the energy gap. 

\begin{figure}[h]
\includegraphics[width=0.9\columnwidth]{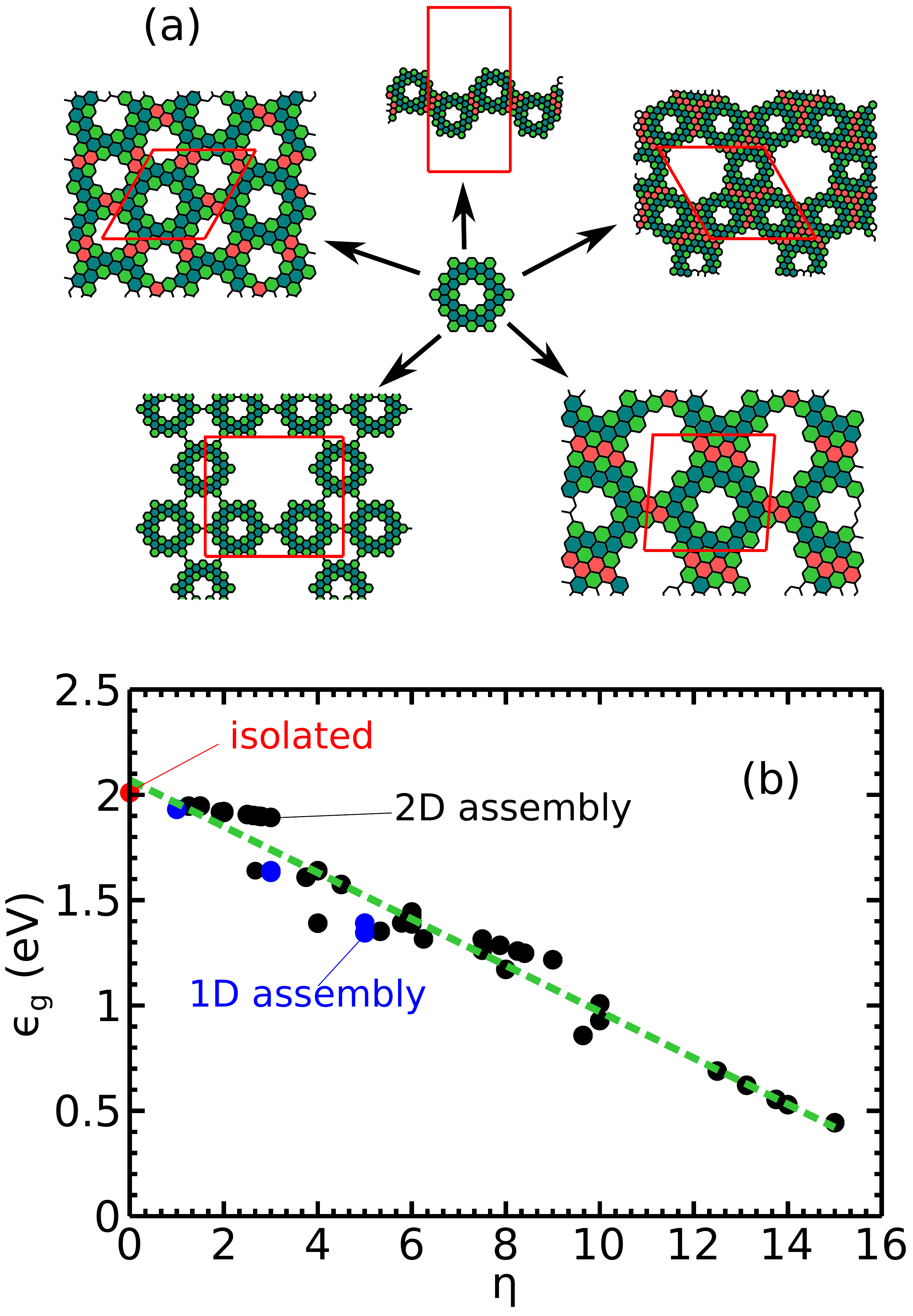}
\caption{\label{gap} (a) Examples of different cycloarene assemblies. All explored assemblies are described in the supplemental material. (b) Energy gap evolution with respect to the inter-molecule bond density $\eta$.}
\end{figure}

From (i) and (ii), we are motivated to extend the cycloarene assemblies to other geometries and track the gap evolution of the system as a function of the inter-molecule bond, Fig.~\ref{gap}(a). We have explored $42$ different assemblies based on 2D lattices \cite{PCCPcrasto2019, PRBcrasto2020, NScrasto2021}; ranging from one molecule per unit cell ($108$ atoms/unit cell) up to $13$ molecules per unit cell ($1404$ atoms/unit cell), depicted in the supplemental material. The energy gap follows a linear tendency with the carbon bonding density ($\eta$) between adjacent cycloarenes, Fig.~\ref{gap}(b). As previously stated for the 1D assemblies, the energy gap's linear behavior is weakly dependent on the molecular assembly geometry. That is, taking different geometry assemblies but with the same $\eta$ presents a low energy gap variation of $\sigma = \sqrt{\langle {\epsilon_g^2} \rangle  - \langle {\epsilon_g} \rangle^2 } \sim 0.05$\,eV. We could extrapolate such linearity to an energy gap curve evolution of $\epsilon = \epsilon_{g}^0 - \alpha \eta = (2.07 - 0.11\,\eta)$\,eV. The $\epsilon_{g}^0$ value from the linear extrapolation (isolated molecule gap) deviates by only $0.06$\,eV from the calculated one. Although the energy gap underestimation in DFT calculations, the energy gap trend is correctly described. As we will show next, the linear factor $\alpha$ originates from the inter-molecule interaction energy, to which DFT offers excellent estimations. {To give a better estimation of the cycloarene assemblies energy gap, we have performed hybrid-functional calculations for the isolated molecule and the three reference assemblies, Fig.~\ref{2d-assembly}(a1)-(d1). Here we found that the bandgap increases by $34\%$ on average, $29\%$ for the isolated molecule, giving an energy gap of $\epsilon_{g}^{0} = 2.48$\,eV. These results are in accordance with the predicted linear gap evolution.} By adjusting $\epsilon_{g}^0$ to the experimental value, a realistic estimation of the assembly range of the bandgap is obtained. {From voltage-dependent differential conductance measurements the experimental gap of cycloarene was estimated to be $2.97$\,eV \cite{JACSfan2020}, representing an increase of $19\%$ in relation to the HSE calculations. Assuming such correction to the other structures, we can expect the cycloarene assemblies bandgap to evolve in the range of [2.97,\,0.81]\,eV. Therefore, the cycloarene assemblies allow tuning porous graphene energy gap in an interval of energy close to the silicon values ($\sim 1$\,eV).}

We can understand the linear behavior of the energy gap with a simple two-level interacting model with hamiltonian
\begin{equation}
H=\begin{pmatrix}
0 & \Delta \\
\Delta & 0
\end{pmatrix},
\end{equation}
to which eigenvalues, for $\Delta$ real, is given by $\epsilon_{\pm} = \pm |\Delta|$, and the energy gap between the two levels $\epsilon_g = \epsilon_{+} - \epsilon_{-} = 2|\Delta|$. This hamiltonian's base $\left\{ |\pm \rangle \right\}$ can always arise in semiconducting/insulating systems with particle-hole (chiral) symmetry. They relate with the HOMO, LUMO states by $|\pm \rangle = \left( |{\rm HOMO} \rangle \pm |{\rm LUMO} \rangle \right)/\sqrt{2}$. By forming an assembly of this molecular system, inter-cell interactions ($V$) will appear in the hamiltonian
\begin{equation}
H_{A} = \begin{pmatrix}
0 & \Delta + V \\
\Delta + V^* & 0
\end{pmatrix},
\end{equation}
which will modify the energy gap to $\epsilon_g = 2\sqrt{(\Delta + V^*)(\Delta + V)}$. The inter-cell coupling ($V$) can be written generally as $V=\delta f(\vec{k})$. Here, $\delta \in \mathbb{R}$ is the coupling strength, and $f(\vec{k})$ is an oscillatory term given by the system periodicity. For simplicity taking a 1D system and $f(k) \in \mathbb{R}$, it assumes the form $f(k)=\cos(k)$. The energy gap at each k point is
\begin{equation}
\epsilon_g = 2|\Delta + \delta \cos(k)|.
\end{equation}
The minimum gap will occur in the $f(k)$ minimum,  $[f(k)]_{\rm min} = -1$,
\begin{equation}
[\epsilon_g]_{\rm min} = 2|\Delta - \delta|.
\end{equation}
Therefore, the energy gap behaves linearly with the inter-cell coupling strength. Similar arguments are possible for 2D assemblies, and a low coupling limit $\delta \ll \Delta$ is straightforwardly obtained for a general case (3D), all obeying the same linear equation. 

\begin{figure}
\includegraphics[width=\columnwidth]{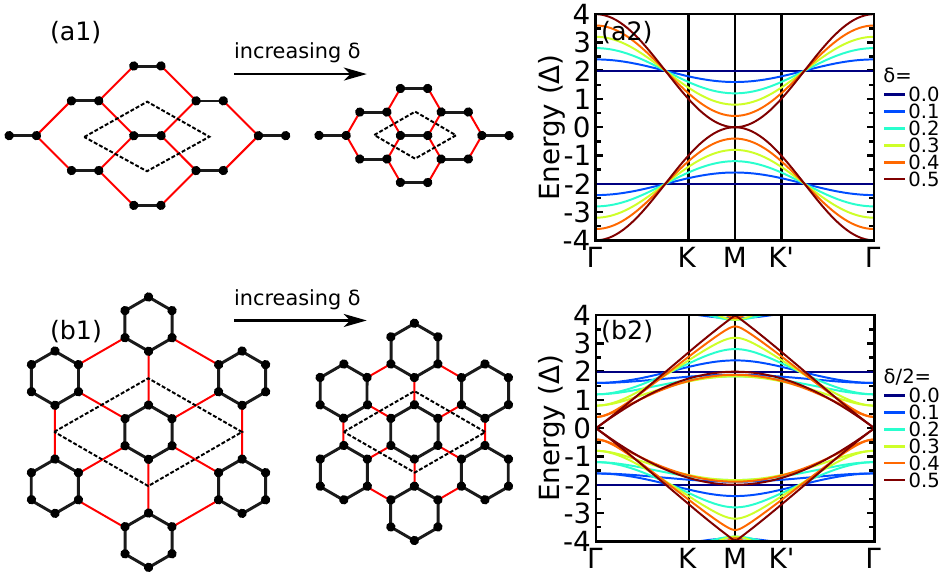}
\caption{\label{model} Model for the gap evolution with inter-cell coupling $\delta$. (a) C$_2$ molecule toward graphene assembly and (b) benzene molecule towards graphene assembly. (a1)-(b1) Show the structure and the $\delta$ coupling in red, and (a2)-(b2) the evolution of the band structure for different $\delta$.}
\end{figure}

The proposed model captures the gap evolution of the studied cycloarene system. Here the number of bonds between neighboring cycloarene rules the inter-molecule coupling, $\delta = (0.11\,\eta)$\,eV. The generality of this model can predict the gap evolution of other non-magnetic carbon-based molecules. For instance, we can exemplify by discussing two different TB models of graphene assembly. In Fig.~\ref{model}(a1), we considered the graphene formed by $C_2$ molecules assembled towards increasing intermolecular interaction. The system gap evolves linearly with the intermolecular interaction, Fig.~\ref{model}(a2). In this illustrative model, the ${|\pm \rangle}$ base states assume precisely the $|A\rangle $ and $|B \rangle$ sublattice states of the $C_2$ molecule. Additionally, we can consider graphene as the assembly of benzene molecules [Fig.~\ref{model}(b1)], where the gap also evolves linearly with $\delta$ [Fig.~\ref{model}(b2)]. This second case is a limit of the well-known Kekule distortion of graphene \cite{NATPHYgutierrez2016}.

{Besides the band gap, the mobility of a semiconductor is key for constructing efficient devices. Organic semiconductors present usually a mobility in the range of $10^{-1}$ to $10^1$\,cm$^{2}$V$^{-1}$s$^{-1}$ \cite{JACSyavuz2015}, while  values around $10^2$\,cm$^{2}$V$^{-1}$s$^{-1}$ are required for practical application [cite]. Graphene is predicted to have a phonon-limited ultrahigh-mobility of $2\,10^{5}$\,cm$^{2}$V$^{-1}$s$^{-1}$ \cite{NATMATgeim2007} with its zero-gap Dirac bands. In order to estimate the electronic mobility in the semiconducting nanoporous graphene we have computed the phonon-limited mobility based on the deformation potential \cite{PRbardeen1950}, which was successfully used to interpret mobility in 2D systems based on the Takagi model \cite{IEEETEDtakagi1994, NATCOMMqiao2014, AEMfang2019}. The mobility can be written as
\begin{equation}
\mu_j = \frac{e {\hslash}^3 C_{2D}}{k_B T m_j^* m_d (E^i)^2},
\end{equation}
with $e$, $\hslash$ and $k_b$, the electron charge, reduced plank constant and Boltzmann constant, respectively; $C_{2D}$ and $m_j^*$ are the elastic modulus and effective mass along the $j$ propagation direction, with $m_d = \sqrt{m_x^* m_{y}^*}$ being the average effective mass. The term $E^i$  is the deformation potential constant for the conduction band minimum (for electron mobility). We have computed $C_{2D}$ and $E^i$ considering a dilatation/compression with a step of $0.5\%$ of the lattice constant. These DFT calculations were performed in the three systems used as reference, Fig~\ref{2d-assembly}(b)-(d). Here we estimate the mobility along the zig-zag direction ($\mu_{zz}$) of each lattice, the values are presented in the Table~\ref{tab1}.

\begin{table}
\small \centering
\caption{\label{tab1} Electron mobility for the three reference structures shown in Fig~\ref{2d-assembly}(b)-(d) along the zig-zag direction.}
\begin{tabular}{cc}
\hline
$\eta$ & $\mu_{zz}$ [cm$^2$V$^{-1}$s$^{-1}$]  \\
\hline
3  &  6.29$\times 10^2$ \\
9  &  3.98$\times 10^3$ \\
15 &  9.99$\times 10^4$ \\
\hline
\end{tabular}
\end{table}

The estimated values are in accordance with 3D nanoporous graphene systems \cite{AMtanabe2016} and show that for suitable assemblies with higher inter-molecular bond-density ($\eta$) the mobility can reach ultra-high values towards the graphene predicted mobility. Particularly, the system with higher $\mu$, which is the experimentally achieved assembly \cite{JACSfan2020} of Fig~\ref{2d-assembly}(d1), present an estimated energy gap of $0.8$\,eV.}

\section{Conclusions}

In summary, we have explored cycloarene self-assemblies as a route to manipulation of the energy gap in graphene-like systems. We have established a rule for the energy gap engineering based on the inter-molecule bond density. The energy gap evolves linearly with the inter-molecule bond density. {That leads to a possible variation of the bandgap from $2.9$\,eV (lower bond density) down to $0.8$\,eV (higher bond density) in cycloarene assemblies, estimated from the experimental energy gap.} We explain the origin of such inter-molecule bond dependence of the energy gap as the $\pi$ bandwidth variation in {the} assemblies. This interpretation allows us to define an effective model for the gap evolution{. Our effective model can be generally applied to predict and understand the bandgap evolution in other non-magnetic carbon molecules, capturing the bandgap in Kekule distorted graphene. Within the cycloarene assemblies, we have shown that for the higher inter-molecular bond density systems (lower bandgap), the phonon-limited electron mobility can reach ultrahigh values up to $10^{4}$\,cm$^2$V$^{-1}$s$^{-1}$.}

\section*{Conflicts of interest}
There are no conflicts to declare.

\section*{Acknowledgements}
The authors acknowledge financial support from the Brazilian agencies FAPESP  (grants 19/20857-0 and 17/02317-2), INCT-Nanomateriais de Carbono , and  National Laboratory for Scientific Computing (LNCC) for providing HPC resources of the SDumont supercomputer.



\balance


\bibliography{bib} 
\bibliographystyle{rsc} 

\end{document}